\documentclass[12pt]{iopart}
\usepackage{iopams}
\usepackage{amssymb,mathptm,latexsym}

\def\be{\begin{equation}}
\def\ee{\end{equation}}
\def\bea{\begin{eqnarray}}
\def\eea{\end{eqnarray}}
\def\ba{\begin{array}}
\def\ea{\end{array}}
\def\nn{\nonumber}
\def\p{\partial}

\begin{document}


\title[Anomalies, effective action and Hawking radiation]{Anomalies, effective action and Hawking
temperatures of a Schwarzschild black hole in the isotropic coordinates}
\author{Shuang-Qing Wu$^{1,2)}$, Jun-Jin Peng$^{1)}$ and Zhan-Yue Zhao}
\address{
$^{1)}$College of Physical Science and Technology, Central China Normal
University, Wuhan, Hubei 430079, People's Republic of China \\
$^{2)}$Institute of Particle Physics, Hua-Zhong Normal University, Wuhan, Hubei 430079,
People's Republic of China} 
\ead{sqwu@phy.ccnu.edu.cn} 

\date{today}

\begin{abstract}
Motivated by the universality of Hawking radiation and that of the anomaly cancellation technique
as well as that of the effective action method, we investigate the Hawking radiation of a Schwarzschild
black hole in the isotropic coordinates via the cancellation of gravitational anomaly. After performing
a dimensional reduction from the four-dimensional isotropic Schwarzschild metric, we show that this
reduction procedure will, in general, result in two classes of two-dimensional effective metrics: the
conformal equivalent and the inequivalent ones. For the physically equivalent class, the two-dimensional
effective metric displays such a distinct feature that the determinant is not equal to the unity ($\sqrt{-g}
\neq 1$), but also vanishes at the horizon, the latter of which possibly invalidates the anomaly analysis
there. Nevertheless, in this paper we adopt the effective action method to prove that the consistent
energy-momentum tensor $T_{~t}^r$ is divergent on the horizon but $\sqrt{-g} T_{~t}^r$ remains finite
there. Meanwhile, through an explicit calculation we show that the covariant energy-momentum tensor
$\widetilde{T}_{~t}^r$ equals to zero at the horizon. Therefore the validity of the covariant regularity
condition that demands that $\widetilde{T}_{~t}^r = 0$ at the horizon has been justified, indicating that
the gravitational anomaly analysis can be safely extrapolated to the case where the metric determinant
vanishes at the horizon. It is then demonstrated that for the physically equivalent reduced metric, both
methods can give the correct Hawking temperature of the isotropic Schwarzschild black hole, while for the
inequivalent one with the determinant $\sqrt{-g} = 1$ it can only give half of the correct temperature.
We further exclude the latter undesired result by taking into account the general covariance of the
energy-momentum tensor under the isotropic coordinate transformation.

\end{abstract}

\submitto{\CQG}


\maketitle


\section{Introduction}

Hawking radiation \cite{SWH} is one of the most important theoretical discoveries in black hole physics,
named after the surname of famous physicist Steven Hawking. The Hawking effect shows that a black hole
is not really black, but can radiate thermally like a black body. Precisely speaking, Hawking radiation
is the quantum effect of field in a classically curved background space-time with a future event horizon.
It has a feature that the thermal radiation is determined universally by the properties of the horizon.
There are several derivations of Hawking radiation. The original one presented by Hawking \cite{SWH} when
he quantized the scalar field theory in a static Schwarzschild space-time is undoubtedly the most direct
method through which Hawking straightforwardly calculated the Bogoliubov coefficients between in- and
out-states of fields in a black hole background. This derivation, however, is very intricate and difficult
to be generalized to more general cases. Later on, other attempts were made now and then.

Recently, there are two popular and relatively simple methods owing to Wilczek \cite{PW,RW,IUW},
attracting a lot of attention \cite{RBH,OthPs,JWC,WPW,BR,KMU,RecPs,BK,WZ,PWu,ZY,BKG}. One is
the semi-classical tunnelling picture proposed by Parikh and Wilczek \cite{PW}, and the other is
the anomaly cancellation method advocated by Robinson and Wilczek (RW) \cite{RW,IUW}. In the former
method, Hawking radiation is visualized as a tunnelling process from the horizon and can be derived
by calculating the semi-classical WKB amplitudes for classically forbidden paths. In the latter one,
Hawking radiation can be understood as a compensating energy-momentum flux required to cancel the
consistent gravitational anomaly at the horizon in order to preserve general covariance at the
quantum level. Obviously, the RW's method is very universal since this kind of derivation of
Hawking radiation via the viewpoint of the anomaly cancellation is only dependent on the anomaly
taking place at the horizon.

In the further development, Iso \textit{et al}. \cite{IUW} extended the RW's method to investigate
Hawking radiation in the case of a charged black hole, by considering gauge anomaly in addition to
gravitational anomaly. In their work, the condition for consistent gauge and gravitational anomaly
cancellation and regularity requirement of covariant anomalies at the horizon, together with the
energy-momentum conservation law, determines Hawking fluxes of the charge and energy-momentum tensor.
Subsequently, the treatment was further generalized \cite{RBH} to the cases of rotating and charged
black holes, where the rotation essentially plays the role of a $SO(2) \simeq U(1)$ gauge field.
Since then, a lot of application \cite{OthPs,JWC,WPW,BR,KMU,RecPs} of this method appeared and were
devoted to investigating Hawking radiation of other different black objects in various dimensions, and
all the obtained results demonstrated that the gauge current and energy-momentum tensor flux, required
to cancel the consistent gauge and gravitational anomalies at the horizon, are exactly equal to that of
Hawking radiation. These results once again show that Hawking radiation is universal and only depends
on the property of the horizon. It appears that these studies successfully support that the RW's
prescription is very universal. However, it should be pointed out that almost all of these efforts were
limited to the case of the determinant $\sqrt{-g} = 1$ of the effective diagonal metric in two dimensions.
Based on the original work of \cite{RW,IUW}, the anomaly cancellation method was extended \cite{WPW}
to the most general case of two-dimensional non-extremal black hole metrics where $\sqrt{-g} \neq 1$.
Nevertheless, the anomaly analysis presented in \cite{WPW} had overlooked a peculiar case of which the
determinant $\sqrt{-g}$ vanishes at the horizon.

It should be emphasized that the original anomaly cancellation method proposed in Refs. \cite{RW,IUW}
encompasses not only the consistent anomaly but also the covariant one. Quite recently, Banerjee and
Kulkarni \cite{BK} suggested that this model can be further simplified by using only the covariant gauge
and gravitational anomalies to derive Hawking radiation from charged black holes. Their simplification
makes the anomaly analysis more economical and conceptually cleaner. An extension of their work to the
case where the determinant $\sqrt{-g} \neq 1$ was done in \cite{WZ}. Based upon these developments, some
direct applications soon appeared in \cite{PWu,ZY}.

Very recently, Banerjee and Kulkarni \cite{BKG} advocated to use the covariant boundary condition and
the covariant energy-momentum tensor to directly calculate the Hawking fluxes from charged black holes
via the effective action method. The results obtained by them are in accordance with those obtained by
the RW's method. They also established the connection of this approach with the calculations based on
the Unruh vacuum. However, the equivalence of the effective action method and the covariant anomaly
method needs to be further clarified in the very general case where $\sqrt{-g} \neq 1$.

Note however, in fact, that there is a subtlety in the anomaly analysis, namely, the validity of the
RW's method implicitly assumes the regular behaviors of the dilaton field and the metric determinant
at the horizon. In other words, the metric determinant should not vanish or diverge on the horizon. This
issue had been neglected in the previous researches since almost all of them only deal with the trivial
case where $\sqrt{-g} = 1$. As will be demonstrated in this paper, the above assumption is apparently
violated in the case for a Schwarzschild black hole in the isotropic coordinate system. Therefore it is
necessary to investigate extensively the most general case where $\sqrt{-g} \neq 1$, in particular, the
special case when $\sqrt{-g}$ vanishes at the horizon. The Schwarzschild black hole in the isotropic
coordinates just provides such an example that its metric determinant vanishes at the horizon.

In this paper, we attempt to apply the gravitational anomaly method and the effective action method
to study Hawking radiation of the Schwarzschild black hole in isotropic coordinate system. Although
the same question had been previously studied \cite{RW,BKG} in the standard Schwarzschild coordinates,
the isotropic coordinates display such a distinct feature that the metric determinant of the reduced
two-dimensional effective and physically equivalent metric is not equal to the unity ($\sqrt{-g} \neq 1$),
but also vanishes at the horizon, the latter of which possibly invalidates the anomaly analysis there
when one applies it to investigate the Hawking radiation in such a coordinate system. The interest to the
isotropic coordinates is that they play an important role in finding a large classes of supersymmetric
black hole solutions in string theory. For example, the famous five-dimensional BMPV black hole solution
\cite{BMPV} can be expressed in terms of isotropic coordinates. Our motivation for this investigation
is also inspired from the fact that Hawking radiation is a universal quantum phenomenon and should not
depend on the concrete choice of different coordinates, but it really relies on the property of the
horizon. Correspondingly, the Hawking temperature derived in different coordinates should be given by
the same value, and not be related with a concrete coordinate system. Our aim is intended to verify
that the RW's method and the effective action method are unrelated with a different choice of coordinates.

After performing a dimensional reduction procedure from 4 to 2 dimensions, we show that the dimensional
reduction technique will, in general, result in two classes of two-dimensional effective metrics: the
equivalent and the inequivalent ones. We define the physically equivalent reduced metrics as those that
neither change the rank of the zeros of the components $g_{tt}$ and $g^{rr}$ of the original four-dimensional
metric, nor flip their signs. Otherwise we shall call them the in-equivalent classes. For the physically
equivalent class, the effective two-dimensional metric displays such a distinct feature that the metric
determinant is not equal to the unity ($\sqrt{-g} \neq 1$), but also vanishes at the horizon, the latter
of which possibly gives rise to a doubt about the reliability of a direct application of the anomaly
analysis there.

In order to prove that the anomaly analysis can be safely extrapolated to the situation when the metric
determinant vanishes at the horizon, we derive the consistent energy-momentum tensor given by the effective
action \cite{RBH,BKG,2defac,BF} in the Unruh vacuum state. We find that the energy-momentum tensor $T_{~t}^r$
is finite at spatial infinity but diverges on the horizon, therefore $\sqrt{-g}T_{~t}^r$ rather than $T_{~t}^r$
itself is regular on the horizon because $\sqrt{-g}$ vanishes there. We verify the regularity condition via
explicitly calculating the covariant energy-momentum tensor and showing that both $\widetilde{T}_{~t}^r$
and $\sqrt{-g}\widetilde{T}_{~t}^r$ vanish at the horizon. This means that the validity of the anomaly
analysis at the horizon can be guaranteed and the RW's method can be safely extrapolated to the case when
$\sqrt{-g}$ vanishes at the horizon. We further demonstrate that for the physically equivalent reduced
metric, both the effective action method and the RW's method can derive the correct Hawking temperature
of the Schwarzschild black hole in the isotropic coordinates, which is exactly in accordance with Hawking's
original result; while for the inequivalent one with $\sqrt{-g} = 1$ it can only give a temperature with
its value just being one-half of the correct one originally derived by Hawking. The latter result is
undesired and can be further excluded by taking into account the covariance of the energy-momentum
tensor under the isotropic coordinate transformation.

Our paper is outlined as follows. In section \ref{sbhics}, we derive the line element of a Schwarzschild
black hole in the isotropic coordinates from the standard coordinates by means of the isotropic coordinate
transformation and calculate its surface gravity via the standard formulae. We compare the difference
between these two coordinate systems in detail. In section \ref{DiRe}, we present the dimensional reduction
of the massless scalar field theory in the isotropic Schwarzschild space-time and demonstrate that this
reduction procedure will, in general, result in two classes of two-dimensional effective metrics: the
equivalent and the inequivalent ones. In section \ref{efacuv}, we compute the consistent energy-momentum
tensor given by the effective action \cite{RBH,BKG,2defac,BF} in the Unruh vacuum state and show that
$T_{~t}^r$ diverges on the horizon, but $\sqrt{-g}T_{~t}^r$ is finite there. Meanwhile, the covariant
energy-momentum tensor is explicitly calculated and shown to be equal to zero at the horizon, verifying
the validity of the covariant regularity condition $\widetilde{T}_{~t}^r = 0$ at the horizon. This fact
confirms us that the validity of the gravitational anomaly analysis at the horizon is guaranteed and the
RW's method can be safely extrapolated to the case where the metric determinant vanishes at the horizon.
Section \ref{cahr} is devoted to deriving the Hawking temperature via the cancellation of the consistent
or covariant anomaly at the horizon. It is shown that for the physically equivalent reduced metric, in
which the resultant determinant vanishes at the horizon, both the effective action method and the anomaly
cancellation method can give a temperature of the isotropic Schwarzschild black hole, which is just the
correct one derived by Hawking in his original work; while for the inequivalent class, with which the
determinant $\sqrt{-g} = 1$ in the two-dimensional effective background field theory, both methods can
only derive a temperature that is one-half of the original one given by Hawking. Finally we further exclude
this incorrect result by taking into account the general covariance of the energy-momentum tensor. Our
results show that the Hawking temperature have the same value in different coordinate systems. In addition,
the equivalence between the effective action method and the RW's method has also been established. The
last section \ref{sumdis} ends up with a summary of our main results and presents our conclusions.

\section{The Schwarzschild black hole: the isotropic
coordinates versus the standard ones } \label{sbhics}

There are several different forms for the metric of a Schwarzschild black hole. Among them, the most popular
one is expressed in the standard Schwarzschild coordinate system as follows
\be
ds^2 = -\Big(1 -\frac{2M}{\hat{r}}\Big)dt^2 +\frac{d\hat{r}^2}{1 -2M/\hat{r}} +\hat{r}^2d\Omega^2 \, ,
\label{sbh}
\ee
where $M$ is the Arnowitt-Deser-Misner (ADM) mass of the black hole, $d\Omega^2$ is the metric on the unit
two-sphere. It is well known that an event horizon of the black hole is determined by $g^{\hat{r}\hat{r}}
= 0$ while a Killing horizon is given by $g_{tt} = 0$. Obviously, the Schwarzschild black hole has an event
horizon that coincides with the Killing horizon at $\hat{r}_H = 2M$ where the metric becomes singular.

For the question in which we interested here, we shall consider the Schwarzschild black hole in the
isotropic coordinates. After performing a coordinate transformation
\be
\hat{r} = r\big(1 +M/2r\big)^2 \, ,
\label{ict}
\ee
the line element (\ref{sbh}) then becomes
\be
ds^2 = -\frac{(1 -M/2r)^2}{(1 +M/2r)^2}dt^2 +\Big(1 +\frac{M}{2r}\Big)^4\big(dr^2 +r^2d\Omega^2\big) \, ,
\label{ibh}
\ee
for which the metric determinant is
\be
\sqrt{-g_4} = r^2\Big(1 -\frac{M}{2r}\Big)\Big(1 +\frac{M}{2r}\Big)^5\sin\theta \, .
\ee
In this isotropic coordinate system, the black hole now has only a Killing horizon which is located at
$r_H = M/2$, corresponding to $\hat{r}_H = 2M$ in the standard coordinate system. We can see that the
metric (\ref{ibh}) is regular at $r_H = M/2$. Now we want to calculate the temperature of the back hole
in the isotropic coordinates. The surface gravity can be computed in the standard manner and is given by
\be
\kappa = -\frac{1}{2}\sqrt{\frac{g^{rr}}{-g_{tt}}}\frac{\p g_{tt}}{\p r}\Big|_{r = r_H} = \frac{1}{4M} \, ,
\ee
so the corresponding Hawking temperature is
\be
T = \frac{\kappa}{2\pi} = \frac{1}{8\pi M} \, ,
\ee
which is exactly the same one as that calculated in the standard coordinate system. Therefore, one can
draw a conclusion that Hawking temperature should be determined by the properties of the horizon but
unrelated with the choice of different coordinates. It is anticipated that an analogous universality
can apply to the effective action method and the gravitational anomaly method.

There are some common characters shared by the black hole space-time in both metric forms, which can be
summarized as follows. Both line elements are asymptotically flat at spatial infinity; The parameter
$M$ is the ADM mass in both metrics; and the surface gravity has the same expression $\kappa = 1/(4M)$
in both coordinate systems.

However, there exist a lot of explicit differences between these two coordinates in differen aspects such
as, the type of horizons, properties of the singularity and the metric determinant, the zeros and signs of
the temporal and radial metric components, etc. In the standard coordinates, the black hole horizon $\hat{r}_H
= 2M$ is both an event horizon and a Killing horizon; the determinant $\sqrt{-\hat{g}_4} = \hat{r}^2\sin\theta$
is regular everywhere; the zeros of the components $g_{tt}$ and $g^{\hat{r}\hat{r}}$ are the same, both metric
components flip their signs at the horizon. By contrast, in the isotropic coordinates, the location $r_H
= M/2$ is only a Killing horizon; the determinant $\sqrt{-g_4} = \sqrt{-\hat{g}_4}\zeta(r)$ vanishes at the
horizon; the zeros of the components $g_{tt}$ and $g^{rr}$ are different, and their signs do not change at
$r_H = M/2$. These differences can be attributed to the fact that the isotropic coordinate transformation
(\ref{ict}) has changed the order of zeros of the metric components and contributed to the metric determinant
a Jacobi factor
\be
\zeta(r) \equiv \frac{\p \hat{r}}{\p r} = 1 -\frac{M^2}{4r^2} \, .
\ee

Finally, the asymptotic behaviors of the tortoise coordinates defined in both systems are different. In the
standard Schwarzschild coordinates, the tortoise coordinate is defined as
\be
\hat{r}_* = \int\frac{\hat{r}~d\hat{r}}{\hat{r} -2M} \simeq 2M\ln (\hat{r} -2M)
 \equiv \frac{1}{2\kappa}\ln (\hat{r} -\hat{r}_H) \, ,
\ee
while it is defined in the isotropic coordinates via
\be
r_* = \int\big(1 +M/2r\big)^3\frac{r}{r -M/2}dr
\simeq 4M\ln (r -M/2) \equiv \frac{1}{2\kappa}\ln (r -r_H)^2 \, .
\ee
Their difference can be seen more clearly via finding the inverse transformation of Eq. (\ref{ict}), namely
\be
2r = \hat{r} -M +\sqrt{\hat{r}^2 -2M\hat{r}} \, ,
\ee
from which we get
\be
\sqrt{\hat{r}^2 -2M\hat{r}} = r -M^2/4r \, , \quad \hat{r} - M = r +M^2/4r \, .
\ee
The above coordinate transformation maps the exterior region $\hat{r} \geq 2M$ to the domain $r \geq M/2$.
Near the horizon $\hat{r}_H = 2M$ (or $r_H = M/2$), the asymptotic behaviors of these two coordinates are
related by
\be
\big(r -M/2\big)^2 \simeq M\big(\hat{r} -2M\big)/2 \, .
\ee

In addition, it is necessary to emphasize that the rank of the singularity (namely, the order of zeros of
the metric component $g_{tt}$) has been changed under the isotropic coordinate transformation. That is, a
simple singular point in the standard Schwarzschild coordinate system now becomes a rank-two singular point
in the isotropic coordinate system.

\section{Dimensional reduction: the equivalent
versus the in-equivalent classes} \label{DiRe}

Since the anomaly cancellation method proposed in Ref. \cite{RW} involves a dimensional reduction of
the space-time from higher dimensions to two dimensions, therefore it is crucial to obtaining a reduced
two-dimensional effective metric that inherits most fundamental properties of the higher-dimensional
one. In this section, we shall show that the dimensional reduction procedure will yield two classes of
two-dimensional effective background metrics: the conformal equivalent and the inequivalent ones. We
present our definition for the equivalent and the inequivalent metrics, and demonstrate that even for
the physically equivalent classes, the reduced two-dimensional effective metrics are not uniquely
determined in the process of dimensional reduction, since they can differ by a regular conformal factor.

Before adopting the RW's method and the effective action method to investigate Hawking radiation from the
isotropic Schwarzschild black hole, we now perform a dimension reduction in the isotropic coordinate system.
Considering a massless scalar field in the background metric (\ref{ibh}), the action can be written as
\bea
S[\varphi] &=& -\frac{1}{2}\int d^4x\sqrt{-g}g^{\mu\nu}\p_{\mu}\varphi\p_{\nu}\varphi
 = \frac{1}{2}\int d^4x\sqrt{-g}\varphi\Box\varphi \nn \\
&=& \frac{1}{2}\int dtdrd\theta d\phi ~r^2\sin\theta ~\varphi\Big\{
 -\frac{(1 +M/2r)^7}{1 -M/2r}\p_t^2 \nn \\
&& +\frac{1}{r^2}\p_r\Big[\Big(r^2 -\frac{M^2}{4}\Big)\p_r\Big]
 +\frac{1}{r^2}\Big(1 -\frac{M^2}{4r^2}\Big)\Delta_{\Omega}\Big\}\varphi \nn \\
&=& \frac{1}{2}\sum_{lm}\int dtdr \varphi_{lm}\Big\{
 -\frac{r^2(1 +M/2r)^7}{1 -M/2r}\p_t^2 +\p_r\Big[\Big(r^2 -\frac{M^2}{4}\Big)\p_r\Big] \nn \\
&& +l(l +1)\Big(1 -\frac{M^2}{4r^2}\Big)\Big\}\varphi_{lm} \, .
\eea
where $\Delta_{\Omega}$ is the angular Laplace operator on the unit two-sphere, with the eigenvalue $l(l+1)$.
In the last line, we have performed a partial wave decomposition $\varphi = \sum_{lm}\varphi_{lm}(t, r)
Y_{lm}(\theta, \phi)$ in terms of the spherical harmonic functions.

Keeping the dominant terms in the vicinity of the horizon ($r_H = M/2$), the action is then simplified as
\be
S[\varphi] \simeq \frac{1}{2}\sum_{lm}\int dtdr ~r^2\varphi_{lm}\Big[
 -\frac{(1 +M/2r)^7}{1 -M/2r}\p_t^2 +\Big(1 -\frac{M^2}{4r^2}\Big)\p_r^2 \Big]\varphi_{lm} \, .
\label{ea1}
\ee
The above near-horizon limit suggests that the action can be effectively replaced by the following one
\be
S[\varphi] \equiv \frac{1}{2}\sum_{lm} \int dtdr \Psi\varphi_{lm}
 \Big[-\frac{1}{\sqrt{fh}}\p_t^2 +\sqrt{fh}\p_r^2\Big]\varphi_{lm} \, ,
\label{ea2}
\ee
in a two-dimensional space-time with the effective metric
\be
ds^2 = -f(r)dt^2 +h(r)^{-1}dr^2 \, .
\label{2dem}
\ee
This indicates that the scalar field theory in the vicinity of the event horizon of the original
four-dimensional black hole space-time (\ref{ibh}) can be effectively described by an infinite
collection of massless scalar fields in the two-dimensional effective background metric (\ref{2dem}).

Comparing Eqs. (\ref{ea1}) with (\ref{ea2}), we get
\be
\frac{\Psi}{\sqrt{fh}} = r^2\frac{(1 +M/2r)^7}{1 -M/2r} \, , \quad
 \Psi\sqrt{fh} = r^2 -\frac{M^2}{4} \, .
\ee
From these equations, we can solve the dilaton factor as $\Psi = r^2(1 +M/2r)^4$, and obtain
\be
\sqrt{fh} = \frac{1 -M/2r}{(1 +M/2r)^3} \, .
\ee
With the help of the metric determinant $\sqrt{-g} = \sqrt{f/h} = \xi(r)$, $f(r)$ and $h(r)$ can be
obtained as follows
\be
f = \frac{1 -M/2r}{(1 +M/2r)^3}\xi(r) \, , \quad  h = \frac{1 -M/2r}{(1 +M/2r)^3\xi(r)} \, .
\ee
Thus the reduced two-dimensional effective metric can be written as a conformal form as
\be
ds^2 = \xi(r)\Big[-\frac{1 -M/2r}{(1 +M/2r)^3}dt^2 +\frac{(1 +M/2r)^3}{1 -M/2r}dr^2\Big] \, ,
\label{crm}
\ee
with $\sqrt{-g} = \xi(r)$ acting as a conformal factor.

At this stage, we have only set the dilaton field as a regular function and left the metric determinant to
be determined. Obviously, this implies that the resultant two-dimensional effective metrics are not unique.
Thus, the dimension reduction procedure leaves us a room to choose many different expressions for the
conformal factors, which can lead to several different results.

It is now a position to determine the concrete expression for the metric determinant $\sqrt{-g} = \xi(r)$.
Note that its general expression assumes a splitting form
\be
\sqrt{-g} = \xi(r) = \big(1 -M/2r\big)^n\chi(r) \, ,
\ee
where the power $n$ is an integer, and $\chi(r)$ is an arbitrary regular function satisfying
\be
\chi(r = r_H) \neq 0 \, , \infty \, ;  \qquad \chi(r\to\infty) = 1 \, .
\label{cond}
\ee
The asymptotic behavior of the determinant at the horizon is mainly controlled by the power $n$. The latter
condition means that $\sqrt{-g} = 1$ when $r\to\infty$. The reason for this assumption will be explained
later.

Now we propose our definition for the equivalent and the inequivalent classes of the reduced two-dimensional
effective metrics.

\textit{Proposition}: The physically equivalent classes for the reduced two-dimensional effective metrics
are those that neither change the zeros of the components $g_{tt}$ and $g^{rr}$ of the original space-time
metric (\ref{ibh}), nor flip their signs at the horizon. Otherwise they shall be called the in-equivalent
classes \footnote{In the case of a rotating black hole, the same question should be addressed in the
dragging coordinate system. See for example, the first reference in \cite{PWu}.} The equivalent effective
metrics are also conformal equivalent, up to a regular conformal factor $\chi(r)$ that satisfies the
conditions (\ref{cond}).

According to the above definition, we can classify the reduced two-dimensional metrics (\ref{crm}) into
two types:

\textbf{(I) Conformal equivalent metrics}: To ensure the physical equivalence of the reduced two-dimensional
metric with the original four-dimensional one, the zeros and signs of the $g_{tt}$ and $g^{rr}$ components of
the metric (\ref{ibh}) should not be changed in the process of dimensional reduction. To retain the zeros and
signs of the components $g_{tt}$ and $g^{rr}$ of the original metric (\ref{ibh}), one must take $n = 1$ for
these equivalent classes. Thus the line element is
\be
ds^2 = \chi(r)\Big[-\frac{(1 -M/2r)^2}{(1 +M/2r)^3}dt^2 +(1 +M/2r)^3dr^2\Big] \, ,
\label{emc}
\ee
with the determinant $\sqrt{-g} = \chi(r)[1 -M/(2r)]$ in which the regular function $\chi(r)$ acts like
a conformal factor.

At this moment, the metric (\ref{emc}) is still not completely determined since $\chi(r)$ can be an arbitrary
regular function as long as it satisfies the conditions (\ref{cond}). As such, the resultant two-dimensional
effective metrics for the equivalent classes are not unique in the process of dimensional reduction, but are
conformal equivalent under a regular conformal transformation. In other words, they are merely determined,
up to a regular conformal factor.

We now explain the reason why one should demand that $\sqrt{-g} = 1$ when $r\to\infty$. This is because
the surface gravity $\kappa$ and $\sqrt{-g} T_{~t}^r$ are conformal invariant physical quantities in two
dimensions, so one must demand $\chi(r\to\infty) = 1$ to obtain consistent physical results of the Hawking
fluxes under a regular conformal transformation. In addition, this condition is also compatible with the
asymptotically flatness of the metric at spatial infinity.

There are a lot of regular functions for $\chi(r)$ that satisfy the conditions (\ref{cond}). One natural
choice is $\chi(r) = 1 +M/(2r)$ so that $\sqrt{-g} = \zeta(r) = 1 -M^2/(4r^2)$, then the line element
becomes
\be
ds^2 = -\frac{(1 -M/2r)^2}{(1 +M/2r)^2}dt^2 +\Big(1 +\frac{M}{2r}\Big)^4dr^2 \, ,
\label{fhrt}
\ee
which is exactly the $(t-r)$-sector of the original line element (\ref{ibh}).

Alternatively, one can, for example, choose $\chi(r) = [1 +M/(2r)]^{-3}$ so that the reduced metric is
\bea
ds^2 = -\frac{(1 -M/2r)^2}{(1 +M/2r)^6}dt^2 +dr^2 \, . \nn
\eea
This line element differs from (\ref{fhrt}) by a regular conformal factor $[1 +M/(2r)]^4$. So even for
the physically equivalent classes, the two-dimensional effective metrics, which are reduced from the
four-dimensional space-time, need not to be the $(t-r)$-sector of the latter. This is different from
the observation \cite{RW}, where the dimension reduction keeps the form of the $(t-r)$-sector unchanged.

It should be noted that the above two metrics are conformal equivalent, both of them give the consistent
results in the anomaly analysis. However in both cases, $\sqrt{-g} = 0$ at $r_H = M/2$, $\sqrt{-g} = 1$
at spatial infinity, and $\sqrt{-g} \neq 1$ elsewhere. That the metric determinant vanishes at the horizon
will give rise to a doubt about the reliability of the anomaly analysis there.

Most related to the RW's method is to compute the surface gravity of the resultant two-dimensional
effective metric so as to reproduce the Hawking temperature of the original four-dimensional black hole.
For the equivalent classes of the reduced two-dimensional metrics, the above two line elements differ
by a regular conformal factor. Their surface gravities can be computed by making use of the standard
method and are found to be the same value as that presented in the four-dimensional case, so the Hawking
temperature is
\be
T = \frac{\kappa}{2\pi} = \frac{1}{4\pi}\sqrt{h/f} \p_r f\big|_{r_H} = \frac{1}{8\pi M} \, ,
\ee
which is just the standard value originally calculated by Hawking.

As will be seen later, both the RW's method and the effective action method can also derive the same
correct Hawking temperature for the conformal equivalent two-dimensional metrics. However, a question
that arises in the anomaly analysis is whether it is still valid for the case when $\sqrt{-g} = 0$ at
$r_H = M/2$. In the next two sections, we will pay our attention to show that the anomaly analysis can
be safely extrapolated to the situation when the metric determinant vanishes at the horizon. A key
observation is to show that $\sqrt{-g}T_{~t}^r$ rather than $T_{~t}^r$ itself is finite on the horizon
and $\widetilde{T}_{~t}^r = 0$ at the horizon.

\textbf{(II) In-equivalent metrics}: The metrics for the in-equivalent classes are those all that obey
$n \neq 1$. In this case, one can arbitrarily select the conformal factor as long as it does not take
$n = 1$. For example, one can simply choose $\sqrt{-g} = \xi(r) = 1$ so that
\be
ds^2 = -\frac{1 -M/2r}{(1 +M/2r)^3}dt^2 +\frac{(1 +M/2r)^3}{1 -M/2r}dr^2 \, .
\label{iem}
\ee
With this choice of the conformal factor, the metric determinant is well-behaved everywhere, in particular,
it is regular on the horizon. This feature is especially suitable to do the analysis via the anomaly
cancellation method.

On the other hand, one can also select the line element to be that of the optical space
\bea
ds^2 = -dt^2 +\frac{(1 +M/2r)^6}{(1 -M/2r)^2}dr^2 \, , \nn
\eea
for which the determinant will diverge on the horizon. Besides, other choice of the conformal factor is
also possible. But at this moment, we only restrict ourself to the two cases listed above.

The temperature for the metric (\ref{iem}) with $f = h = [1 -M/(2r)]/[1 +M/(2r)]^3$ is
\be
T = \frac{\kappa}{2\pi} = \frac{1}{4\pi}\p_rf|_{r_H} = \frac{1}{16\pi M} \, ,
\ee
which is one-half of the value of Hawking's original result. On the other hand, the temperature calculated
for the optical space is null, a rather different result. The same undesired result can also derived via
the anomaly cancellation technique and the effective action method. So it turns out that the in-equivalent
classes for the reduced metrics are physical meaningless, since all of them can not correctly reproduce the
standard value for the Hawking temperature of a Schwarzschild black hole. Therefore, these types of the
reduced metrics are undesirable in our discussions made below.

Before ending up with this section, let us briefly summarize the main results obtained in this section. We
have generally shown that the original four-dimensional space-time (\ref{ibh}) can be effectively reduced
to two typically different kinds of two-dimensional metrics: the equivalent and the in-equivalent ones.
For the physically equivalent two-dimensional metrics, we propose that the process of dimension reduction
should make no change in the rank of the singularity of the original space-time metric, otherwise the
resultant metrics belong to the in-equivalent classes. Note that the differences between the two-dimensional
line elements (\ref{fhrt}) and (\ref{iem}) resemble much like those in the four-dimensional case listed in
the last section. It should be emphasized that even for the equivalent classes, the reduced two-dimensional
effective metrics need not to be completely identical with, but conformal to the $(t-r)$-sector of the
original four-dimensional isotropic line element, up to a regular conformal factor. Only for the conformal
equivalent classes can one derive the same correct temperature as that given by Hawking.

The above issue have not been exploited in the previous literature. In fact, the same technical question
also exists in the dimensional reduction of the four-dimensional Schwarzschild black hole in the standard
coordinates. To see this point more clearly, let us consider the near-horizon limit of the action of a
massless scalar field
\be
S[\varphi] \simeq \frac{1}{2}\sum_{lm}\int dtd\hat{r} ~\hat{r}^2\varphi_{lm}\Big[
 \frac{-1}{1 -2M/\hat{r}}\p_t^2 +\Big(1 -\frac{2M}{\hat{r}}\Big)\p_{\hat{r}}^2 \Big]\varphi_{lm} \, ,
\ee
which suggests it can be replaced by an effective action in the two-dimensional reduced metric $d\hat{s}^2
= \sqrt{-\hat{g}}ds^2$, where $\sqrt{-\hat{g}}$ acts as a conformal factor and the line element
\be
ds^2 = -\Big(1 -\frac{2M}{\hat{r}}\Big)dt^2 +\frac{d\hat{r}^2}{1 -2M/\hat{r}} \, ,
\label{rmsc}
\ee
is the $(t-r)$-sector of the four-dimensional Schwarzschild space-time in the standard coordinates \cite{IUW}.

For the physically equivalent classes, one must impose that $\sqrt{-\hat{g}} \neq 0$ at $\hat{r}_H = 2M$
and $\sqrt{-\hat{g}} = 1$ when $\hat{r}\to\infty$. The reduced two-dimensional metrics are not unique but are
conformal equivalent. They need not to be in exact agreement with the $(t-r)$-sector of the four-dimensional
metric, but a simple choice for the conformal factor is $\sqrt{-\hat{g}} = 1$, which can reach to this end
\cite{RW}. In addition, it is easy to find that the two-dimensional effective metric (\ref{fhrt}) can
be obtained from (\ref{rmsc}) via the isotropic coordinate transformation (\ref{ict}), while the metric
(\ref{iem}) can not.

In the remaining sections, we will demonstrate that in the case of a physically equivalent metric, both
the effective action method and the gravitational anomaly cancellation method can give a temperature, which
is exactly the same as the Hawking's original one. But for the in-equivalent metrics, the RW' method can not
derive the correct Hawking temperature. For instance, in the case of a two-dimensional metric (\ref{iem})
with $\sqrt{-g} = 1$, both methods give a temperature which is only one-half of the Hawking's standard one.
In the remainder of this paper, we shall only focus on the case of a physically equivalent metric for which
we must deal with the situation when the metric determinant vanishes at the horizon.

\section{Effective action, Unruh
vacuum and Hawking flux}\label{efacuv}

In the preceding section, we have obtained, by means of the dimensional reduction, the physically equivalent
classes of the resultant two-dimensional metrics which are conformal, by a regular conformal factor, to the
$(t-r)$-sector of the four-dimensional Schwarzschild black hole in the isotropic coordinates.

From now on, we shall base upon our discussions in terms of a two-dimensional effective metric (\ref{fhrt})
written in the general form (\ref{2dem}) as
\bea
ds^2 = -f(r)dt^2 +h(r)^{-1}dr^2 \, , \nn
\eea
where
\bea
f = \frac{(1 -M/2r)^2}{(1 +M/2r)^2} \, , \qquad  h = \frac{1}{(1 +M/2r)^4} \, . \nn
\eea
The meter determinant is given by
\be
\sqrt{-g} = \sqrt{f/h} = 1 -\frac{M^2}{4r^2} \, .
\ee
It is obvious that $\sqrt{-g} = 0$ at $r = r_H = M/2$, $\sqrt{-g} = 1$ as $r\to\infty$, and $\sqrt{-g}
\neq 1$ elsewhere.

For later convenience, we denote the surface gravity at any point $r$ as
\be
K = \frac{1}{2}\sqrt{h/f}f^{\prime} = \frac{Mr^2}{(r +M/2)^4} \, ,
\ee
where a prime ($\prime \equiv \p_r$) denotes the derivative with respect to the radial coordinate, here
and hereafter. We also introduce the antisymmetric tensors $\varepsilon^{\mu\nu} = \epsilon^{\mu\nu}/
\sqrt{-g}$ and $\varepsilon_{\mu\nu} = \sqrt{-g}\epsilon_{\mu\nu}$ in terms of the antisymmetric tensor
density $\epsilon^{\mu\nu}$ with the convention $\epsilon^{tr} = \epsilon_{rt} = 1$.

The scalar curvature in the background (\ref{2dem}) reads
\be
R = -2\sqrt{h/f}K^{\prime} = \frac{4Mr^3}{(r +M/2)^6} \, .
\ee
Note that both $K$ and $R$ vanish at spatial infinity, and remain finite at the horizon ($r_H = M/2$).
The surface gravity at the horizon is given by $\kappa = \lim_{r\to r_H}K = 1/(4M)$.

As mentioned before, the validity of the RW's method might not be guaranteed at the horizon as the metric
determinant vanishes there. In order to justify the validity of the anomaly analysis at the horizon, we
have to prove that $T_{~t}^r$ diverges on the horizon but $\sqrt{-g}T_{~t}^r$ remains finite there. Our
strategy will be to calculate the anomaly induced stress tensor in the Unruh vacuum by using the trace
anomaly in the context of a two-dimensional black hole with the effective metric (\ref{fhrt}). The main
reason for this is because the boundary condition adopted in the gravitational anomaly method is consistent
with the Unruh vacuum, and the anomaly cancellation method with this boundary condition gives the same
result as the one derived using the trace anomaly method. What is more, we also explicitly compute the
covariant energy-momentum tensor and show that it indeed vanishes at the horizon, thus verifying the
validity of the regularity condition $\widetilde{T}_{~t}^r = 0$ at the horizon.

In the case of a Schwarzschild black hole, there are a lot of work \cite{2defac,BF} to derive Hawking flux
from effective actions \cite{RBH,BKG,2defac,BF} in the two-dimensional effective background. Although the
two-dimensional effective theory contains a dilaton field, the effect of dilaton does not change the property
of $(r, t)$-component of the energy-momentum tensor and accordingly the Hawking flux is independent of the
dilaton background. Once we impose the boundary condition, the value of the energy flux is determined only
by the value of anomalies at the horizon. Therefore in the following, it will suffice for us to make use of
the effective action to calculate the energy flux in the two-dimensional effective metric without dilaton
backgrounds.

\subsection{Consistent energy-momentum
tensor and Hawking flux}

It is known that the two-dimensional trace anomaly is
\be
T_{~\mu}^\mu = -\frac{R}{24\pi} \, .
\label{2dcta}
\ee
The quantum anomaly induced effective action \cite{RBH,BF} can be obtained by functional integration of
this trace anomaly (\ref{2dcta}). It is the well-known Polyakov-Liouville's non-local action \cite{PL}
\be
\Gamma[g] = -\frac{1}{96\pi}\int d^2x \sqrt{-g}R\Box^{-1}R \, ,
\ee
where $\Box$ is the covariant Laplacian operator in two dimensions.

We recognize this effective action reflects the structure of the conformal anomaly of Eq. (\ref{2dcta}).
This action is non-local, but it can be transformed to a local form by introducing an auxiliary field
$\psi$ \cite{RBH,BF}
\be
\Gamma[\psi,g] = -\frac{1}{96\pi}\int d^2x \sqrt{-g}(-\psi\Box\psi +2\psi R) \, ,
\ee
where the auxiliary field $\psi$ satisfies the equation $\Box\psi = R$.

The quantum anomaly induced stress tensor is defined via the variation of this effective action with
respect to $g^{\mu\nu}$ as follows
\be\hspace*{-1.8cm}
T_{\mu\nu} = \frac{2}{\sqrt{-g}}\frac{\delta \Gamma[\psi, g]}{\delta g^{\mu\nu}}
 = -\frac{1}{48\pi}\Big[\nabla_{\mu}\psi\nabla_{\nu}\psi -2\nabla_{\mu}\nabla_{\nu}\psi
  +g_{\mu\nu}\big(2R -\frac{1}{2}\nabla^{\rho}\psi\nabla_{\rho}\psi\big)\Big] \, .
\ee
This stress tensor obeys the trace anomaly equation (\ref{2dcta}) and the conservation law
\be
T_{~\mu}^\mu = -\frac{R}{24\pi} \, , \qquad \nabla_{\mu}T_{~\nu}^\mu = 0 \, .
\ee

For the two-dimensional effective metric (\ref{2dem}), the equation of motion for the auxiliary field
$\psi$ becomes
\be
-\frac{1}{\sqrt{fh}}\p_t^2\psi +\p_r\big(\sqrt{fh}\p_r\psi\big) = -2\p_r K \, ,
\ee
whose general solution of this equation is \cite{RBH,BF}
\be
\psi = at +\int \frac{b -2K}{\sqrt{fh}} dr = at +br_* -\ln (f) \, ,
\ee
where $a$ and $b$ are constants to be specified by imposing appropriate boundary condition, and $r_*$
is the tortoise coordinate defined by $r_* = \int dr/\sqrt{fh} = r +M^2/(4r) +2M\ln(r) +4M\ln
[1 -M/(2r)] +const$.

Using this solution for $\psi$ and its partial derivatives
\be
\p_t\psi = a \, , \qquad \p_r\psi = (b -2K)/\sqrt{fh} \, ,
\ee
the energy-momentum tensor is calculated as
\bea
T_{tt} &=& -\frac{1}{96\pi}\big(a^2 +b^2 -4K^2 +8\sqrt{fh}K^{\prime}\big) \, , \quad \\
T_{tr} &=& T_{rt} = -\frac{ab}{48\pi\sqrt{fh}} \, , \\
T_{rr} &=& -\frac{1}{96\pi fh}\big(a^2 +b^2 -4K^2\big) \, .
\eea

Introducing the Eddington-Finkelstein (EF) coordinate system $\{u,v\}$, where $u= t -r_*$ and $v = t +r_*$,
all components of the anomaly induced stress tensor in the EF null frame are
\bea
T_{uu} &=& \frac{1}{4}\big(T_{tt} -2\sqrt{fh}T_{tr} +fhT_{rr}\big)
 = -\frac{1}{192\pi}\big[(a -b)^2 -4K^2 +4\sqrt{fh}K^{\prime}\big] \nn \\
 &=& -\frac{1}{192\pi}\big[(a -b)^2 -4\kappa^2\big] \quad~~ (r = r_H) \, , \quad \\
T_{uv} &=& T_{vu} = \frac{1}{4}\big(T_{tt} -fhT_{rr}\big)
 = -\frac{1}{48\pi}\sqrt{fh}K^{\prime} \, , \\
T_{vv} &=& \frac{1}{4}\big(T_{tt} +2\sqrt{fh}T_{tr} +fhT_{rr}\big)
 = -\frac{1}{192\pi}\big[(a +b)^2 -4K^2 +4\sqrt{fh}K^{\prime}\big] \nn \\
 &=& -\frac{1}{192\pi}\big(a +b\big)^2 \quad~~ (r\to\infty)  \, .
\eea

It is known that the gravitational anomaly analysis selects the Unruh vacuum to be the relevant quantum
state. So the appropriate boundary condition for our purpose should correspond to the Unruh vacuum, and
such a boundary condition should be imposed on the auxiliary field $\psi$.

Therefore, to specify the unknown constants $a$ and $b$, we shall adopt the Unruh vacuum as the boundary
condition that assumes there is no ingoing flux (incoming radiation) on past null infinity ($T_{vv} = 0$
as $r\to\infty$) and the energy-momentum tensor has to be regular on the future horizon ($T_{uu} = 0$ for
$r = r_H$). That is
\be
T_{uu} = 0 \quad (r = r_H) \, , \qquad\qquad
T_{vv} = 0 \quad (r\to\infty)  \, .
\ee
From this boundary condition, we obtain $a = -b = \pm\kappa$.

Thus, we get the explicit expression for the $(r, t)$-component of the energy-momentum tensor
\be
T_{~t}^r = -\frac{ab}{48\pi\sqrt{f/h}} = \frac{\kappa^2}{48\pi\sqrt{f/h}} \, ,
\ee
from which, the energy flux of Hawking radiation can be evaluated by taking the asymptotic limit
($r\to\infty$). This yields
\be
T_{~t}^r(r\to\infty) = \frac{\kappa^2}{48\pi} \, .
\label{hremf}
\ee
On the other hand, $T_{~t}^r$ diverges on the horizon ($T_{~t}^r \to\infty$ as $r\to r_H$), but $\sqrt{-g}
T_{~t}^r$ always remains finite everywhere,
\be
\sqrt{-g}T_{~t}^r = \frac{\kappa^2}{48\pi} \, .
\ee
This means that $\sqrt{-g}T_{~t}^r$ is regular at the horizon.

It is quite interesting to examine the behavior of the auxiliary field $\psi$. The asymptotic requirement
$b = -a$ yields $\psi \sim -bu$ as $r\to\infty$, $T_{\mu\nu}$ describes an outgoing radiation; whereas
the regularity condition on the horizon $b = \kappa$ reveals that $\psi \sim -\kappa v$ as $r\to r_H$,
$T_{\mu\nu}$ is finite on the future horizon and describes an ingoing flux of negative energy radiation.
Both $\psi$ and $\nabla_{\mu}\psi$ are regular on the future horizon.

\subsection{Covariant energy-momentum
tensor and Hawking flux}

We now turn to the covariant energy-momentum tensor given by \cite{BKG,HLca}
\be
\widetilde{T}_{~\nu}^{\mu} = -\frac{1}{96\pi}\Big(\frac{1}{2}D^{\mu}\psi D_{\nu}\psi
 -D^{\mu}D_{\nu}\psi +\delta^{\mu}_{~\nu}R\Big) \, ,
\ee
where the auxiliary field $\psi$ still obey the equation $\Box\psi = R$, while $D_{\mu}$ is the chiral
covariant differential operator defined by
\be
D_{\mu} = \nabla_{\mu} +\varepsilon_{\mu\nu}\nabla^{\nu}
 = \varepsilon_{\mu\nu}D^{\nu} \, .
\ee
This covariant energy-momentum tensor simultaneously satisfies the conformal anomaly equation and the
covariant anomaly equation as follows
\be
\widetilde{T}_{~\mu}^\mu = -\frac{R}{48\pi} \, , \qquad
 \nabla_{\mu}\widetilde{T}_{~\nu}^\mu = -\frac{1}{96\pi}\varepsilon_{\mu\nu}\p^{\mu}R \, .
\ee

Inserting the explicit solution for $\psi$ and its chiral covariant derivatives
\be
D_t\psi = -\sqrt{f/h}D^r\psi = a -b +2K \, ,
\ee
the $(r, t)$-component of the covariant energy-momentum tensor can be computed as
\be
\widetilde{T}_{~t}^{r} = \frac{1}{192\pi\sqrt{f/h}}
 \big[(a -b)^2 -4K^2 +4\sqrt{fh}K^{\prime}\big] \, .
\ee
Taking the $r\to\infty$ limit with which $\sqrt{-g}\to 1$, we get the covariant form of the energy flux
\be
\widetilde{T}_{~t}^{r}(r\to\infty) = \frac{1}{192\pi}\big(a -b\big)^2 \, .
\ee
From the requirement that the energy flux at spatial infinity should be $\kappa^2/(48\pi)$, we obtain
$a = b \pm 2\kappa$.

Substituting this condition back to the expression of the covariant stress tensor, we find
\bea
\widetilde{T}_{~t}^{r} &=& \frac{1}{48\pi\sqrt{f/h}}\big(\kappa^2 -K^2 +\sqrt{fh}K^{\prime}\big) \nn \\
 &=& \frac{\kappa^2}{48\pi}\frac{(1 -M/2r)^3}{(1 +M/2r)^9}\Big(1 +\frac{6M}{r}
 +\frac{35M^2}{2r^2} +\frac{3M^3}{2r^3} +\frac{M^4}{16r^4}\Big) \, .
\eea
At the horizon, $\widetilde{T}_{~t}^{r}$ approaches to zero as $r\to r_H$. This fact supports the validity
of the regularity condition proposed in \cite{IUW}, which demands that $\widetilde{T}_{~t}^{r}$ should
vanish at the horizon. It suggests the covariant regularity condition ($\widetilde{T}_{~t}^{r} = 0$ at
$r = r_H$) can be imposed as usual. In other words, the validity of the anomaly analysis at the horizon
can be guaranteed so that the gravitational anomaly cancellation method can be safely extrapolated to the
case when $\sqrt{-g} = 0$ at $r = r_H$.

On the other hand, if we consider the near-horizon behavior
\be\hspace*{-1.8cm}
\sqrt{-g}\widetilde{T}_{~t}^{r} = \frac{1}{192\pi}\big[(a -b)^2 -4K^2 +4\sqrt{fh}K^{\prime}\big]
 = \frac{1}{192\pi}\big[(a -b)^2 -4\kappa^2\big] \quad (r = r_H) \, , \quad
\ee
we can see that $\sqrt{-g}\widetilde{T}_{~t}^{r}$ is apparently well-behaved everywhere. In particular,
the above condition ($a = b \pm 2\kappa$) makes $\sqrt{-g}\widetilde{T}_{~t}^{r} \equiv 0$ become an
\emph{identity} at the horizon. Conversely, if we let $\sqrt{-g}\widetilde{T}_{~t}^{r} = 0$, then we
arrive at $\widetilde{T}_{~t}^{r}(r\to\infty) = \kappa^2/(48\pi)$.

Finally, it should be pointed out that the general expression for the covariant stress tensor
\be
\sqrt{-g}\widetilde{T}_{~t}^{r} = \frac{1}{48\pi}
 \big(\kappa^2 -K^2 +\sqrt{fh}K^{\prime}\big)
 = \Big\{ \ba{ll} 0 &\quad (r = r_H) \\
 \frac{\kappa^2}{48\pi} &\quad (r\to\infty) \ea \, ,
\ee
is identical to the result derived by using the anomaly cancellation method in the next section,
establishing the equivalence of these two methods.

As a comparison with the energy-momentum tensor derived in the last subsection, the components of the
covariant energy-momentum tensor are given by
\bea
\widetilde{T}_{tt} &=& -\frac{1}{48\pi}\big(\kappa^2 -K^2 +2\sqrt{fh}K^{\prime}\big) \, , \\
\widetilde{T}_{tr} &=& \frac{1}{48\pi\sqrt{fh}}\big(\kappa^2 -K^2 +\sqrt{fh}K^{\prime}\big) \, , \\
\widetilde{T}_{rr} &=& -\frac{1}{48\pi fh}\big(\kappa^2 -K^2\big) \, .
\eea
Transformed into the EF null frame, they are
\bea
\widetilde{T}_{uu} &=& -\frac{1}{48\pi}\big(\kappa^2 -K^2 +\sqrt{fh}K^{\prime}\big)
 = -\frac{\kappa^2}{48\pi} \quad~~ (r\to\infty) \, , \\
\widetilde{T}_{uv} &=& -\frac{1}{96\pi}\sqrt{fh}K^{\prime} \, , \\
\widetilde{T}_{vv} &=& 0 \, .
\eea

\subsection{Comparison with results
in the standard coordinates}

We now compare the results obtained in the above two subsections with those derived in the two-dimensional
standard Schwarzschild coordinates (\ref{rmsc}) with the effective metric given by \cite{IUW}
\bea
ds^2 = -\Big(1 -\frac{2M}{\hat{r}}\Big)dt^2 +\frac{d\hat{r}^2}{1 -2M/\hat{r}} \, , \nn
\eea
with $f = h = 1 -2M/\hat{r}$. At the horizon ($\hat{r}_H = 2M$), the surface gravity is $\kappa = 1/(4M)$.

The previous analysis can be naively applied to this case. Now we have $\sqrt{-\hat{g}} = 1$ and
\be
K = \frac{1}{2}\p_{\hat{r}}f = \frac{M}{\hat{r}^2} \, , \qquad
R = -2\p_{\hat{r}}K = \frac{4M}{\hat{r}^3} \, .
\ee
Note also that the tortoise coordinate is replaced by $\hat{r}_* = \int d\hat{r}/f = \hat{r} +2M\ln
(\hat{r} -2M) +const$.

Then the consistent and covariant energy-momentum tensors are, respectively, given by \cite{RBH,BKG}
\bea
T_{~t}^{\hat{r}} &=& \frac{\kappa^2}{48\pi} \, , \\
 \widetilde{T}_{~t}^{\hat{r}} &=& \frac{1}{48\pi}\Big(\kappa^2 -K^2 -\frac{1}{2}fR\Big) \nn \\
 &=& \frac{\kappa^2}{48\pi} \Big(1 +\frac{4M}{\hat{r}} +\frac{12M^2}{\hat{r}^2}\Big)
 \Big(1 -\frac{2M}{\hat{r}}\Big)^2 \nn \\
 &=& \Big\{ \ba{ll} 0 &\quad (\hat{r} = \hat{r}_H) \, , \\
 \frac{\kappa^2}{48\pi} &\quad (\hat{r}\to\infty) \, . \ea
\eea
This means that the validity of the covariant regularity boundary condition $\widetilde{T}_{~t}^{\hat{r}}
= 0$ at the horizon is justified by the effective action method \cite{RBH,BKG,2defac,BF} in the case of
the two-dimensional Schwarzschild black hole (\ref{rmsc}).

In the above analysis, we have obtained a correct Hawking temperature $T = 1/(8\pi M)$ for the effective
metric (\ref{fhrt}). On the contrary, if we directly apply the effective action method to the case of
the in-equivalently reduced metric (\ref{iem}), we will get a temperature $T = 1/(16\pi M)$, which is half
of the correct one originally derived by Hawking. The same incorrect temperature can also be deduced by
the RW's method. This unexpected result can be further excluded by taking into account the covariance of
the energy-momentum tensor under the isotropic coordinate transformation.

As mentioned before, the two-dimensional effective metric (\ref{fhrt}) can be obtained from the one
(\ref{rmsc}) in the standard coordinates via the isotropic coordinate transformation (\ref{ict}), while
the two-dimensional line element (\ref{iem}) can not. The general covariance implies that the result
derived from the latter metric (\ref{iem}) is non-physical.

Under this isotropic coordinate transformation $\hat{r} = r[1 +M/(2r)]^2$, the consistent and covariant
energy-momentum tensors transform as
\bea
T_{~t}^{\hat{r}} &=& \frac{\p \hat{r}}{\p r}T_{~t}^{r}
 = \Big(1 -\frac{M^2}{4r^2}\Big)T_{~t}^{r} \, , \\
 \widetilde{T}_{~t}^{\hat{r}} &=& \frac{\p \hat{r}}{\p r}\widetilde{T}_{~t}^{r}
 = \Big(1 -\frac{M^2}{4r^2}\Big)\widetilde{T}_{~t}^{r} \, .
\eea
Using $\sqrt{-g} = 1 -M^2/(4r^2)$ and noting $\sqrt{-\hat{g}} = 1$, we find that
\bea
T_{~t}^{\hat{r}} &=& \sqrt{-g}T_{~t}^{r} = \frac{\kappa^2}{48\pi} \, , \\
 \widetilde{T}_{~t}^{\hat{r}} &=& \sqrt{-g}\widetilde{T}_{~t}^{r}
 = \frac{1}{48\pi}\Big(\kappa^2 -K^2 -\frac{1}{2}fR\Big) \, , \quad
\eea
which means both $\sqrt{-\hat{g}}T_{~t}^{\hat{r}} = \sqrt{-g}T_{~t}^{r}$ and $\sqrt{-\hat{g}}\widetilde
{T}_{~t}^{\hat{r}} = \sqrt{-g}\widetilde{T}_{~t}^{r}$ are not only covariant but also conformal invariant
in two dimensions.

The regularity boundary condition \cite{IUW} at the horizon naturally leads to
\be
\sqrt{-g}\widetilde{T}_{~t}^{r} = \widetilde{T}_{~t}^{\hat{r}} = 0 \, ,
\ee
which is in accordance with our previous result. This fact reflects the covariance and the universality of
the effective action method.

To end up with this section, our main results can be summarized in one word, that is, $\sqrt{-g}T_{~t}^{r}$
is regular everywhere and the regularity boundary condition $\widetilde{T}_{~t}^{\hat{r}} = 0$ at the horizon
has been obviously verified. This conclusion suggests that the gravitational anomaly analysis can work well
at the horizon.

\section{Anomalies and Hawking temperatures}\label{cahr}

In this section, we will include the RW' method extended in \cite{WPW} for self-consistence. The general
case considered in \cite{WPW} for the two-dimensional black hole is non-extremal, including the extremal
black hole as a special subcase. It is of particular interest to consider the metric of the two-dimensional
Schwarzschild black hole in the isotropic coordinates. In this case, the metric determinant $\sqrt{-g} = 0$
at $r = r_H$, $\sqrt{-g} = 1$ as $r\to\infty$, and $\sqrt{-g} \neq 1$ elsewhere. Our previous analysis carried
through in \cite{WPW} had obviously overlooked this case.

Because the metric determinant equals to zero at the horizon, the gravitational anomaly method will become
problematic there. Therefore a careful investigation is needed for the reliability of the anomaly analysis
at the horizon. However, we have argued in the preceding section that $\sqrt{-g}T_{~t}^{r}$ remains finite
and $\widetilde{T}_{~t}^{\hat{r}}$ is indeed equal to zero at the horizon, which ensures the covariant
regularity boundary condition $\widetilde{T}_{~t}^{\hat{r}} = 0$ at the horizon can be imposed as usual.
Therefore, the anomaly analysis is reasonable at the horizon, and the RW's method is applicable to the case
considered here. In the subsequent analysis, we will apply the anomaly cancellation method and the effective
action method to derive the Hawking temperature in the background space-time (\ref{iem}) recast into the
general form (\ref{2dem}).

\subsection{Consistent anomaly and
covariant boundary condition}

The gravitational anomaly method proposed in \cite{RW} interprets Hawking radiation as a compensating energy
flux that cancels the consistent or covariant gravitational anomaly at the horizon. The main idea goes as
follows. Near the horizon, when omitting the ingoing modes that can not affect the physics outside of the
horizon at classical level, the effective two-dimensional field theory becomes chiral and exhibits an anomaly.
Therefore the quantum effective action breaks down the symmetry under the diffeomorphism transformation,
which contradicts the fact that the underlying theory is covariant. As a result, in order to preserve the
covariance of theory at quantum level, Hawking radiation must be included to cancel the anomaly near the
horizon. In this method, the contribution from the dilaton can be neglected thanks to the static space-time.

In the following, our starting point for the consistent anomaly analysis will be based upon the general
effective metric (\ref{2dem}). Let us consider the effective field theory in the two-dimensional background
space-time (\ref{2dem}).  We divide the region outside of the horizon into two parts: the near-horizon
region $[r_H, r_H +\varepsilon]$ and the far-away region $[r_H +\varepsilon, +\infty)$, where $\varepsilon$
will be taken the $\varepsilon \to 0$ limit ultimately. In the region $[r_H, r_H +\varepsilon]$, the theory
becomes chiral after neglecting the classically irrelevant ingoing modes and the energy-momentum tensor there
satisfies the anomaly equation.

It is well known that the minimal form of the consistent gravitational anomaly for right-handed fields reads
\cite{AWBK}
\be
\nabla_{\mu}T_{~\nu}^{\mu} = \frac{1}{96\pi\sqrt{-g}}
 \epsilon^{\beta\alpha}\p_{\alpha}\p_{\mu}\Gamma_{\nu\beta}^{\mu}
 = \frac{1}{\sqrt{-g}}\p_{\mu}N_{~\nu}^{\mu} \, ,
\label{mcga}
\ee
while the covariant anomaly takes the form
\be
\nabla_{\mu}\widetilde{T}_{~\nu}^{\mu}
 = \frac{-1}{96\pi}\sqrt{-g}\epsilon_{\mu\nu}\p^{\mu}R
 = \frac{1}{\sqrt{-g}}\p_{\mu}\widetilde{N}_{~\nu}^{\mu} \, ,
\ee
where $\epsilon^{\mu\nu}$ is the antisymmetric tensor density with the upper and the lower case
$\epsilon^{tr} = \epsilon_{rt} = 1$.

Solving the anomaly equations in terms of the effective background metric (\ref{2dem}), we get the
$(r, t)$-components of $N_{~\nu}^{\mu}$ and $\widetilde{N}_{~\nu}^{\mu}$
\bea
N^r_{~t} &=& \frac{1}{192\pi}\big(hf^{\prime\prime} +f^{\prime}h^{\prime}\big) \, , \\
\widetilde{N}^r_{~t} &=& \frac{1}{96\pi}\Big(hf^{\prime\prime}
 +\frac{1}{2}f^{\prime}h^{\prime} -\frac{h}{f}f^{\prime 2}\Big)
 = \frac{1}{48\pi}\big(\sqrt{fh}K^{\prime} -K^2\big) \, .
\eea

Introducing two scalar (top hat and step) functions $\Theta(r) = \Theta(r -r_H -\varepsilon)$ and $H(r) = 1
-\Theta(r)$, the total energy-momentum tensor can be expressed as
\be
T_{~\nu}^{\mu} = T_{(O)\nu}^{\mu}\Theta(r) +T_{(H)\nu}^{\mu}H(r) \, ,
\ee
where $T_{(O)\nu}^{\mu}$ is the covariantly conserved stress tensor satisfying $\nabla_{\mu} T_{(O)\nu}^{\mu}
= 0$, and $T_{(H)\nu}^{\mu}$ must obey Eq. (\ref{mcga}). Integrating both equations, we get
\bea
\sqrt{-g}T_{(O)t}^r &=& a_O \, , \\
\sqrt{-g}T_{(H)t}^r &=& a_H +N_{~t}^r -N_{~t}^r(r_H) \, ,
\eea
where $a_O$ and $a_H$ are two integration constants, and $a_O$ is the value of the energy-momentum tensor
flux at spatial infinity, representing the observed Hawking radiation from the horizon. Note that both
$T_{(O)t}^r$ and $T_{(H)t}^r$ diverges on the horizon because $\sqrt{-g} = 0$ there, but $\sqrt{-g}
T_{(O)t}^r$ and $\sqrt{-g}T_{(H)t}^r$ remain finite at the horizon.

Our main task is to fix $a_O$. In order to do so, let us consider the $\nu = t$ component of the
conservation law
\be\hspace*{-1.8cm}
\sqrt{-g}\nabla_{\mu}T_{~t}^{\mu} = \p_r\big[\sqrt{-g}T_{~t}^r\big]
 = \p_r\big[N_{~t}^rH(r)\big] +\big[N_{~t}^r
 +\sqrt{-g}\big(T_{(O)t}^r -T_{(H)t}^r\big)\big]\delta(r -r_H) \, .
\ee
The first term can be cancelled by the quantum effect of the ingoing modes. One can make the energy-momentum
tensor free of anomaly by re-defining a new one via $\bar{T}_{~t}^r = T_{~t}^r -N_{~t}^rH(r)/\sqrt{-g}$.
Therefore, in order to keep the covariance of the underlying theory, we only need to make the coefficient
of the second term disappear at the horizon
\be
\sqrt{-g}\big(T_{(O)t}^r -T_{(H)t}^r\big)\big|_{r_H} +N_{~t}^r(r_H) = 0 \, ,
\label{constr}
\ee
that is to say,
\be
a_O = a_H -N_{~t}^r(r_H) \label{eqa} \, .
\ee
This equation can not fix $a_O$ completely unless the value of the constant $a_H$ is known.

In Ref. \cite{IUW}, a covariant regularity condition is imposed to demand that the covariant energy-momentum
tensor should vanish at the horizon ($\widetilde{T}_{~t}^r = 0$ at $r = r_H$). Now that we have explicitly
shown in the last section for the case considered here, $\widetilde{T}_{~t}^r$ is indeed equal to zero at
the horizon, the same boundary condition still can be imposed so as to be capable of applying the RW's
method. As a result, we have $\sqrt{-g} = \widetilde{T}_{~t}^r = 0 = \sqrt{-g}\widetilde{T}_{~t}^r$ at
the horizon.

Since the covariant and consistent anomalies are related by \cite{WPW}
\be
\sqrt{-g}\widetilde{T}_{~t}^r = \sqrt{-g}T_{~t}^r +\widetilde{N}_{~t}^r -N_{~t}^r
 = \sqrt{-g}T_{~t}^r +\frac{h}{192\pi f} \big(ff^{\prime\prime} -2f^{\prime 2}\big)\, ,
\label{ccr}
\ee
as a consequence, the regularity condition yields
\be
a_H = N_{~t}^r(r_H) -\widetilde{N}_{~t}^r(r_H) \, .
\ee
Therefore we get the expression of Hawking flux by taking the asymptotic limit
\be
T_{(O)t}^r(r\to\infty) = a_O = -\widetilde{N}^r_{~t}(r_H) = \frac{\kappa^2}{48\pi} \, ,
\label{hrf}
\ee
which is obtained via the cancellation of the consistent gravitational anomaly at the horizon. It agrees
with the results found in \cite{BK,PWu}.

The expression for the consistent energy-momentum tensor
\be
\sqrt{-g}T_{(O)t}^r = -\widetilde{N}^r_{~t}(r_H) = \frac{\kappa^2}{48\pi} \, ,
\ee
is in exact accordance with the result obtained in the last section. This fact verifies the equivalence
between the RW's method and the effective action method.

We now point out the difference between the above result and the previous one given in \cite{WPW}
\be
a_O = N^r_{~t}(r_H) = \frac{1}{192\pi}\big(hf^{\prime\prime}
 +f^{\prime}h^{\prime}\big)\big|_{r_{H}} \, .
\label{tpw}
\ee
In Ref. \cite{WPW}, we have considered the general non-extremal black hole case in which $f(r_H) = h(r_H)
= 0$. This means that the functions $f(r)$ and $h(r)$ are assumed to have the same asymptotic behaviors,
namely, they simultaneously behave like $(r -r_H)$ or $(r -r_H)^2$ near the horizon. It is not difficult
to find that the relation $a_O = N^r_{~t}(r_H) = -\widetilde{N}^r_{~t}(r_H)$ indeed and only holds true
in such cases.

\subsection{Covariant anomaly only}

Recently, the authors in \cite{BK} suggested it is much cleaner to use only the covariant anomaly to derive
the Hawking flux. In the above subsection, we have adopted the model proposed in \cite{IUW} to calculate
the energy-momentum flux through the cancellation of the consistent anomaly, but to fix it we have imposed
a regular condition that requires $\widetilde{T}_{~t}^r = 0$ at the horizon. This implies that we have
adopted two different anomalies to obtain the Hawking flux. By contrast, the only input advocated in
\cite{BK} is the covariant anomaly. Nevertheless, we can see below that the result is unchanged.

In the exterior region, there is no anomaly and the covariant energy-momentum tensor is conserved,
$\nabla_{\mu}\widetilde{T}^{\mu}_{(O)\nu} = 0$. Near the horizon, the energy-momentum tensor obeys
the covariant anomaly equation, $\sqrt{-g}\nabla_{\mu}\widetilde{T}^{\mu}_{(H)\nu} = \p_{\nu}
\widetilde{N}^{\mu}_{~\nu}$. Solving both equations for the $\nu = t$ component,
\bea
\p_r\big[\sqrt{-g}\widetilde{T}^r_{(O)t}\big] &=& 0 \, , \\
\p_r\big[\sqrt{-g}\widetilde{T}^r_{(H)t}\big] &=& \p_r\widetilde{N}^r_{~t} \, ,
\eea
yields the solutions
\bea
\sqrt{-g}\widetilde{T}^r_{(O)t} &=& \widetilde{a}_O \, , \\
\sqrt{-g}\widetilde{T}^r_{(H)t} &=& \widetilde{a}_H +\widetilde{N}^r_{~t} -\widetilde{N}^r_{~t}(r_H) \, ,
\eea
where $\widetilde{a}_O$ and $\widetilde{a}_H$ are two integration constants.

Writing the covariant stress tensor as a sum of two combinations $\widetilde{T}^{\mu}_{~\nu} =
\widetilde{T}^{\mu}_{(O)\nu}\Theta(r) +\widetilde{T}^{\mu}_{(H)\nu}H(r)$, we find
\be\hspace*{-1.8cm}
\sqrt{-g}\nabla_{\mu}\widetilde{T}^{\mu}_{~t} = \p_r\big[\sqrt{-g}\widetilde{T}^r_{~t}\big]
 = \p_r\big[\widetilde{N}^r_{~t}H(r)\big] +\big[\widetilde{N}^r_{~t}
 +\sqrt{-g}\big(\widetilde{T}^r_{(O)t} -\widetilde{T}^r_{(H)t}\big)\big]\delta(r -r_H) \, .
\ee
The first term has to be cancelled by the quantum effect of the incoming modes. As before, it implies
the existence of a Wess-Zumino term modifying the covariant energy-momentum tensor as $\bar{\widetilde
{T}_{~t}^r} = \widetilde{T}^r_{~t} -\widetilde{N}^r_{~t}H(r)/\sqrt{-g}$ which is anomaly free provided
the coefficient of the last term vanishes. This yields the condition
\be
\widetilde{a}_O = \widetilde{a}_H -\widetilde{N}^r_{~t}(r_H) \, .
\ee

As before, the constant $\widetilde{a}_H$ can be fixed by the covariant regularity boundary condition
$\widetilde{T}_{~t}^r = 0$ at the horizon, which gives $\widetilde{a}_H = 0$. Hence the covariant
energy-momentum tensor flux is given by
\be
\widetilde{T}_{(O)t}^r(r\to\infty) = \widetilde{a}_O
 = -\widetilde{N}^r_{~t}(r_H) = \frac{\kappa^2}{48\pi} \, .
\ee
This is just the energy flux from black body radiation.

Finally, the covariant energy-momentum tensors can be expressed as
\bea
\sqrt{-g}\widetilde{T}^r_{(O)t} &=& -\widetilde{N}^r_{~t}(r_H) = \frac{\kappa^2}{48\pi} \, , \\
\sqrt{-g}\widetilde{T}^r_{(H)t} &=& \widetilde{N}^r_{~t} -\widetilde{N}^r_{~t}(r_H)
 = \frac{1}{48\pi}\big(\kappa^2 -K^2 +\sqrt{fh}K^{\prime}\big) \, ,
\eea
which coincides with the results obtained in the last section. This indicates that both the gravitational
anomaly method \cite{RW,IUW,BK} and the effective action method \cite{RBH,BKG,BF} give the same consistent
results. Thus the equivalence between these two approaches is generally justified.

On the other hand, the above anomaly analysis can be directly applied to the effective metric (\ref{rmsc})
in the standard coordinates and the obtained results coincide with those obtained in \cite{RW,IUW,BKG}.
Moreover, we find that $\sqrt{-\hat{g}}T_{~t}^{\hat{r}} = \sqrt{-g}T_{~t}^{r}$ and $\sqrt{-\hat{g}}
\widetilde{T}_{~t}^{\hat{r}} = \sqrt{-g}\widetilde{T}_{~t}^{r}$ are invariant under the isotropic
coordinate transformation. Combined with these results, it can be concluded that both methods are
generally covariant and universal, supporting the universality of the Hawking radiation.

\subsection{Hawking temperatures of the
isotropic Schwarzschild black hole}

With the previous preparation in hand, it is now a position to apply the gravitational anomaly method
and the effective action method to reproduce the Hawking temperature of a Schwarzschild black hole in
the isotropic coordinates. First of all, according to the spirit of the RW's method, in order to cancel
the gravitational anomaly at the horizon so that the general covariance can be restored at the quantum
level, one must identify the Hawking flux with the thermal flux of two-dimensional black body radiation
at a temperature $T = \kappa/(2\pi)$. Next, the effective action method directly adopts the anomaly induced
stress tensors to calculate the energy flux at spatial infinity. As we have seen before, these two methods
are equivalent and give the universal expression for the Hawking flux: $-\widetilde{N}^r_{~t}(r_H) =
\kappa^2/(48\pi)$), which can be identical with the energy flux ($\Phi = \pi T^2/12$) of two-dimensional
black body radiation at the Hawking temperature. This means that we can use the relation $\Phi = -\widetilde
{N}^r_{~t}(r_H)$ to determine the Hawking temperature of the isotropic Schwarzschild black hole.

Now we are ready to explicitly calculate the Hawking temperature of the Schwarzschild black hole in the
isotropic coordinates. As mentioned before, the dimension reduction procedure yields two classes of
two-dimensional effective metrics: the conformal equivalent and the in-equivalent ones. Two typical
representatives for them are given by the two-dimensional metrics (\ref{fhrt}) and (\ref{iem}),
respectively. Both the RW's method and the effective action method are viable and can be applied
to these two cases.

First, let us begin with the equivalent classes. The two-dimensional effective metric (\ref{fhrt}) is just
the $(t-r)$-sector of the original four-dimensional isotropic Schwarzschild space-time. In this case, the
dimension reduction keeps the rank of the singularity unchanged. Since the metric determinant is $\sqrt{-g}
= 1 -M^2/(4r^2)$, it will vanish at the horizon, causing a doubt about the reliability of a direct application
of the RW's method to this case. Using the effective action method, we have obviously verified that $\sqrt{-g}
T_{ ~t }^{r}$ is finite and $\widetilde{T}_{~t}^{r}$ vanishes at the horizon, showing that the covariant
regularity condition $\widetilde{T}_{~t}^{r} = 0$ at the horizon can be imposed as usual. This in turn
indicates that the validity of the anomaly analysis at the horizon can be guaranteed. Applying the RW's
method and the effective action method to this case, we can derive a temperature $T = 1/(8\pi M)$, which
is exactly equal to the correct one originally given by Hawking.

Next, we turn to consider the two-dimensional background metric (\ref{iem}) for the in-equivalent classes.
In this case, we have $f(r) = h(r) = [1 -M/(2r)]/[1 +M/(2r)]^3$ and $\sqrt{-g} = 1$. Both the anomaly
analysis and the effective action method are directly applicable to calculate the Hawking flux. At the
horizon ($r_H = M/2$), due to $f(r_H) = h(r_H) = 0$, the energy flux is given by $\Phi = f^{\prime 2}
(r_H)/(192\pi)$. This will give a temperature $T = 1/(16\pi M)$, which is only one-half of the correct
Hawking temperature. The reason for this is that the rank of the singularity has been changed in the
process of dimension reduction.

Therefore, for the conformal equivalent effective line elemnet(\ref{fhrt}), both methods can reproduce the
correct Hawking temperature, while they can not do so for the in-equivalent reduced line element (\ref{iem}).
The different results for the derived temperatures arise from the different choice of a conformal factor
in the process of dimensional reduction, since it can have an important effect on the order of the zeros
of the component $g_{tt}$ in the resultant two-dimensional effective metric. The discrepancy of the
obtained results serve a technical warning on how to obtain the physically equivalent reduced metric
in the two-dimensional effective theory.

Finally, taking into account the general covariance of the energy-momentum tensor under coordinate
transformations, we can further exclude the undesired result for the Hawking temperature with a $1/2$
factor in the case of the in-equivalent reduced metric (\ref{iem}). The reason for this is because that
the energy-momentum tensor is covariant under the isotropic coordinate transformation (\ref{ict}); both
the effective action method and the RW's method are covariant. Furthermore, the two-dimensional effective
metric (\ref{fhrt}) can be obtained from the one (\ref{rmsc}) in the standard coordinates via the isotropic
coordinate transformation (\ref{ict}), while the two-dimensional line element (\ref{iem}) can not. So the
general covariance implies that only for the conformal equivalent classes can one derive the expected
temperature as that given by Hawking.

\section{Concluding remarks}\label{sumdis}

In this paper, we have utilized the effective action method and the gravitational anomaly method to derive
the Hawking temperature of a Schwarzschild black hole in the isotropic coordinates. Our motivation is
originated from the following facts: First of all, Hawking radiation is a universal quantum phenomenon
which relies merely on the property of the horizon and is unrelated with a concrete coordinate system.
Correspondingly, the Hawking temperature should have a unique value, independent of the concrete choice
of different coordinates. In other words, the Hawking temperature should have the same value although the
metric can be expressed in different forms by adopting different coordinate systems. Next, the RW's method
is universal since the derivation of Hawking radiation via the viewpoint of the anomaly cancellation is
only dependent on the anomaly taking place at the horizon. To preserve general covariance, it requires
an outgoing thermal flux to eliminate the gravitational anomaly at the horizon. One can anticipate the
effective action method to be universal also. Third, both methods are manifestly generally covariant.
Therefore, it is expected that these two methods can deal with Hawking radiation in different coordinate
settings and give the same consistent result. That is, if both methods are successful to derive the
Hawking temperature of a Schwarzschild black hole in the standard coordinate system, the result obtained
by these methods should be unchanged in a different coordinate system (here the isotropic coordinates).

Since both methods work within the two-dimensional effective field theory, so a key trick is how to reduce
the higher-dimensional space-times to two dimensions. The dimension reduction procedure plays a crucial
role in deriving the resultant two-dimensional effective background metrics so that they belong to the
physically equivalent classes.

The Schwarzschild black hole in the isotropic coordinates displays a lot of distinct characters that differ
from the standard one. It also provides an obvious example that one has to deal with the case in which not
only $\sqrt{-g} \neq 1$ but also $\sqrt{-g} = 0$ at the horizon for the conformal equivalent reduced metrics.
Thanks to these intriguing features, it is of particular interest to check the effectiveness of the RW's
method and the effective action method in the isotropic Schwarzschild space-time. Our results have
demonstrated that both methods can derive the correct Hawking temperature for the conformal equivalent
classes of the reduced effective metric (\ref{fhrt}).

The main results obtained in our paper are summarized as follows. (i) The dimensional reduction procedure
yields two classes of typically different metrics. The physically equivalent reduced metrics are defined
as those that neither change the rank of the zeros of the components $g_{tt}$ and $g^{rr}$ of the original
space-time metric, nor flip their signs. These metrics are also conformal equivalent, up to a regular
conformal factor that satisfies the conditions (\ref{cond}). (ii) The equivalence between the effective
action method and the gravitational anomaly method are generally established. (iii) The covariance and the
universality of both methods are obviously justified via the isotropic coordinate transformation. (iv) The
covariant regularity boundary condition imposed in the RW's method has been verified by using the effective
action method to explicitly calculate the covariant energy-momentum tensor, so the validity of RW's method
has also been justified. (v) For the conformal equivalent reduced metric (\ref{fhrt}), both methods can
successfully reproduce the standard value of the Hawking temperature for the isotropic Schwarzschild black
hole. These results support not only the universality of these two methods but also that of Hawking radiation.

Based upon the RW's method and the effective action method, our discussions about the gravitational anomaly
analysis presented in the isotropic coordinates are very general. They have an immediate application to the
stringy black holes since a large classes of black hole solutions found in string theory are expressed
in the isotropic coordinate system, especially for those supersymmetric black holes. It is interesting
to check the effectiveness of these two methods in other different coordinate system case as well.

\section*{Acknowledgements}

This work is partially supported by the Natural Science Foundation of China under Grant No. 10675051.

\section*{References}

\providecommand{\href}[2]{#2}\begingroup\raggedright

\endgroup

\end{document}